\documentclass[aps,prb,twocolumn,floats,epsfig,pdflatex]{revtex4-2}
\usepackage{amsmath}
\usepackage{amsfonts}
\usepackage{graphicx}
\usepackage{amssymb}
\usepackage{amsbsy}
\usepackage{color}
\usepackage{ulem}

\newcommand{\beq}{\begin{equation}}
\newcommand{\eeq}{\end{equation}}
\newcommand{\bea}{\begin{eqnarray}}
\newcommand{\eea}{\end{eqnarray}}

\newcommand{\la}{\langle}
\newcommand{\ra}{\rangle}

\newcommand{\om}{\omega}

\newcommand{\ham}{\mathcal{H}}

\newcommand{\ptl}{\partial}

\newcommand{\cda}{c^{\dagger}}

\begin{document}

\title{Signatures of fragmentation for periodically driven fermions}
\author{Somsubhra Ghosh$^1$, Indranil Paul$^2$, and
K. Sengupta$^1$}
\affiliation{$^1$School of Physical Sciences, Indian Association for the
Cultivation of Science, Kolkata 700032, India. \\
$^2$Universit\'{e} Paris Cit\'{e}, CNRS, Laboratoire Mat\'{e}riaux
et Ph\'{e}nom\`{e}nes Quantiques, 75205 Paris, France.}
\date{\today}

\begin{abstract}

We study the possible signatures of prethermal strong Hilbert space
fragmentation (HSF) for one-dimensional (1D) fermions subjected to a
periodic drive. We extend the results of Phys. Rev. Lett. {\bf 130},
120401 (2023) to show the possibility of such fragmentation for a
large class of experimentally relevant drive protocols. Moreover, we
demonstrate the persistence of HSF when the fermion chain is taken
away from half-filling. Both these analysis indicate the robustness
of the fragmentation phenomenon reported earlier. We also provide an
alternate derivation of the Floquet Hamiltonian of the driven chain
which yields insight into the generic nested commutator structure of
its higher order terms. Finally, we study the density-density
out-of-time-correlators (OTOC) of the driven chain both away and
near the special drive frequencies at which its first order Floquet Hamiltonian 
exhibits fragmentation. We show that these OTOCs, for a chain with open 
boundary condition, exhibit a distinct periodic unscrambling of information at special 
drive frequencies; such unscrambling can therefore serve as a marker of prethermal HSF.
We provide an approximate analytic explanation of the role of HSF
behind such periodic unscrambling and discuss experiments which can detect 
signatures of strong HSF in such driven chains.

\end{abstract}


\maketitle

\section {Introduction}
\label{int}

Closed quantum systems driven out of equilibrium have become
increasingly important subject of research in recent years
\cite{rev1,rev2,rev3}. One of the central questions in this field
pertains to the long-time behavior of local correlation functions of
these systems. In most case, the behavior of such correlation
functions can be understood from the eigenstate thermalization
hypothesis (ETH) \cite{eth1,eth2}. ETH predicts eventual
thermalization, under unitary Hamiltonian dynamics, of a generic
many-body quantum state which can initially be far from equilibrium;
it is one of the central paradigms for understanding long-time
behavior of a generic ergodic many-body system. ETH also holds, with
minor modifications, for periodically driven systems, where the
driven system is ultimately expected to heat up to infinite
temperature \cite{fleth1}.

ETH relies on the ergodicity of a generic quantum system and is
known to fail when it is violated. Such ergodicity violation can
occur in integrable models due to the presence of a large number of
conserved quantities \cite{rev1}. In addition, it fails in the
presence of strong disorder leading to many-body localization and
consequent violation of ergodicity \cite{mblref1,mblref2,mblexp}. A
more subtle and weaker failure of ETH occurs due to emergent
symmetry sectors in an otherwise generic quantum systems. The
presence of such symmetries typically lead to tower of states which
are protected from the other thermal states in its Hilbert space.
Thus any quantum dynamics starting from an initial state which
belongs to this sectors can not thermalize; such states are often
called quantum scars \cite{scarrev1, scar1,scar2,scar3,scar4}. The
violation of ETH in this case is weak since it only happens if the
initial state has large overlap with the states in the scar sector.
The number of such states, for a one-dimensional (1D) system of
length $L$, is typically ${\rm O}(L)$; they form a tiny fraction of
the total number of states in the Hilbert space which is ${\rm
O}(e^L)$. Thus these systems display ETH violating dynamics for a
small fraction of initial states.

Another, recently found, violation of ETH occurs in non-integrable
quantum systems due to the presence of kinetic constraints. The
Hamiltonian of such quantum systems, expressed as a matrix in the
computational basis, breaks down into an exponentially large number
of dynamically disconnected fragments. The presence of such a large
number of disconnected sectors is to be contrasted with those
occurring from the presence of conserved global quantities; the
latter only leads to an ${\rm O}(L)$ disconnected symmetry sectors.
This phenomenon is known as strong Hilbert space fragmentation (HSF)
\cite{fraglit1,fraglit2,fraglit3,fraglit4,fraglit5}. Such a strong
fragmentation naturally breaks ergodicity since any generic initial
state, which belongs to a given fragment, can not, under the action
of the system Hamiltonian, explore states in the Hilbert space that
belong to other fragments. Most of the Hamiltonians studied in this
context are 1D spin or fermionic models
\cite{fraglit1,fraglit2,fraglit3,fraglit4}; however, more recently
some higher-dimensional models have also been shown to exhibit
strong HSF \cite{fraglit5}.

More recently, the generalization of strong HSF to periodically
driven quantum system has been studied \cite{ourprl}. It has been
shown that a periodically driven fermionic chain can show signatures
of Hilbert space fragmentation at special drive frequencies over a
large prethermal timescale; the extent of this prethermal timescale
depends on the drive amplitude and can be quite large in the large
drive amplitude regime. The signatures of such prethermal
fragmentation can be found in entanglement entropy, autocorrelation
function and equal-time correlators of such driven system; each of
these quantities show departure from their counterparts in ergodic
quantum systems \cite{ourprl} demonstrating a clear realization of
prethermal strong HSF. In addition, such prethermal HSF in driven
quantum systems can lead to interesting oscillatory dynamics of
correlators for certain initial states which has no counterpart for
HSF realized in an equilibrium setting \cite{ourprl}.

In this work, we extend the results obtained in Ref.\
\onlinecite{ourprl} in several ways. First, we show that such
signatures of fragmentation can be obtained for a much wider class
of drive protocols. This makes the prethermal fragmentation
phenomenon much more relevant to standard experiments using
ultracold atom platform which we discuss. Second, we provide a
comprehensive analysis of the Floquet Hamiltonian. The analytical
expression for the first order Floquet Hamiltonian, $H_F^{(1)}$,
derived using Floquet perturbation theory (FPT), was presented in
Ref.\ \onlinecite{ourprl}; here we provide an alternate derivation
of the Floquet Hamiltonian up to second order in perturbation
theory. This analysis provides insight into the commutator structure
of the higher order terms that was not apparent from the previous
derivation. It also provides an estimate of the frequency range over
which the first order Floquet Hamiltonian provides a qualitatively
accurate description of the dynamical evolution in the prethermal
regime. Third, we show that the signature of fragmentation persists
when the driven chain is taken away from half-filling. This shows
the robustness of the prethermal fragmentation phenomenon and points
out the possibility of its experimental realization for a wide range
of fermion filling fraction. Finally, we study the density-density
out-of-time correlator (OTOC) for the driven fermion chain. We show
that the behavior of such OTOC is qualitatively different at special
frequencies at which the system exhibits signatures of prethermal HSF. 
In particular, for a driven fermion chain with open boundary condition, we 
find, starting from a  initial ${\mathbb Z}_2$ state, 
unscrambling of information manifested through periodic revival of 
the OTOC. We analyze this phenomenon in details, provide an analytic, albeit qualitative, 
understanding of its mechanism, and tie it to the fragmented 
structure of the first-order Floquet Hamiltonian, $H_F^{(1)}$, of the driven chain 
obtained using FPT. Our results thus demonstrate that 
OTOCs can server as markers for prethermal HSF in a driven system.

The organization of the rest of this work is as follows. In Sec.\ \ref{sec:formalism}, we 
present a derivation of the Floquet Hamiltonian which brings out its nested commutator structure. Next, in Sec.\
\ref{flo}, we discuss the different classes of drive protocols which
allows for signature of prethermal fragmentation and also derive the
higher order Floquet Hamiltonian corresponding to them. This is followed by Sec.\
\ref{ofil} where we demonstrate signature of prethermal HSF away
from half-filling. Next, in Sec.\ \ref{sec:otoc1}, we discuss the
behavior of OTOC in such driven system. Finally, we discuss our main
results and conclude in Sec.\ \ref{dissc}. Some details of the calculation 
are presented in the Appendices.

\section{Formalism}
\label{sec:formalism}

In this section, we outline the derivation of the Floquet Hamiltonian of the 
driven fermion chain. Our derivation brings out the nested commutator structure 
of the Floquet Hamiltonian and also addresses a more general class of drive protocols 
for which the fermion chain exhibits prethermal HSF.  

\subsection{Preliminary}
\label{subsec:preliminary}

Consider a time dependent quantum mechanical system described by the Hamiltonian
\beq
\label{eq:H1}
\ham(t) = \ham_0 (t) + \ham_1,
\eeq
where all the time dependence is in the zeroth order term $\ham_0$. The term $\ham_1$,
which in the following will be treated perturbatively, has no explicit time dependence. 
From Schrodinger equation 
$i \hbar \ptl_t \psi(t) = [\ham_0(t) + \ham_1] \psi(t)$, and the definition of the time 
evolution operator $U(t, 0)$ via $\psi(t) = U(t,0) \psi(0)$, we get
\beq
\label{eq:U1}
i \hbar \frac{\ptl}{\ptl t} U(t, 0) = [\ham_0(t) + \ham_1] U (t, 0).
\eeq

The evaluation of $U(t, 0)$ can be broken into two steps. To do so, we write \cite{fptrev,fpt1,fpt2}
\beq
\label{eq:U2}
U(t, 0) = U_0(t, 0) W(t, 0),
\eeq
where $U_0(t, 0)$ is the exact time evolution operator in the absence of the $\ham_1$ term.
The first step, which is simple, is to evaluate $U_0(t, 0)$ which is given by
\beq
\label{eq:U3}
U_0(t, 0) = \exp[-\frac{i}{\hbar} \int_0^t d \tau \ham_0 (\tau)].
\eeq
In the above the time ordering in front of the exponential can be omitted since the 
operator $\ham_0(\tau)$ at different times commute. The second step, which is non-trivial,
is to compute $W(t, 0)$ that encodes the time evolution due to $\ham_1$. This is performed
perturbatively. Using $i \ptl_t U_0(t, 0) = \ham_0 U_0(t, 0)$ and Eqs.~\eqref{eq:U1} and 
\eqref{eq:U2}, we get
\beq
\label{eq:W1}
i \hbar \frac{\ptl}{\ptl t} W(t, 0) = \ham_p(t) W (t, 0),
\eeq
where
\beq
\label{eq:H2}
\ham_p(t) \equiv U_0(t, 0)^{-1} \ham_1 U_0(t, 0).
\eeq
Using Eq.~\eqref{eq:W1}, the perturbative expansion is
\begin{align}
\label{eq:W2}
W(t, 0) &= 1 - \left(\frac{i}{\hbar}\right) \int_0^t d \tau \ham_p(\tau)  \, +
\nonumber\\
&\left(-\frac{i}{\hbar}\right)^2 \int_0^t d \tau_1 \ham_p(\tau_1) \int_0^{\tau_1} 
d \tau_2 \ham_p(\tau_2)  \, +
\nonumber\\
&\left(-\frac{i}{\hbar}\right)^3 \! \int_0^t \! d \tau_1 \ham_p(\tau_1) \int_0^{\tau_1} \! d \tau_2 
\ham_p(\tau_2) \int_0^{\tau_2} \! d \tau_3 \ham_p(\tau_3)
\nonumber\\
&+ \cdots.
\end{align}
Note, in the above the operator $\ham_p(\tau)$ at different times do not commute.

The above formulation can also be viewed as a series expansion in a rotating frame for the 
following reason. Consider a time dependent unitary transformation $V(t)$ between a 
laboratory to a rotating reference frame with the initial condition $V(0) = 1$. 
The wavefunction in the rotating frame is 
$\psi_r(t) = V^{\dagger} (t) \psi(t)$, and an operator in the same frame is 
$\mathcal{O}_r(t) = V^{\dagger}(t) \mathcal{O} V(t)$, where $\psi(t)$ and $\mathcal{O}$
are the wavefunction and the operator in the laboratory frame, respectively. Simultaneously,
the Hamiltonian $\ham(t)$ in the laboratory frame transforms to $\ham_r(t)$ in the rotating 
frame. By demanding that $i \hbar \ptl_t \psi_r(t) = \ham_r(t) \psi_r(t)$ we get
\beq
\label{eq:H3}
\ham_r(t) = V^{\dagger}(t) \ham(t) V(t) - i \hbar V^{\dagger}(t) \dot{V}(t),
\eeq
where $\dot{V} \equiv \ptl_t V(t)$. Furthermore, if we define the time evolution 
operator $U_r(t_1, t_2)$ in the rotating frame by 
$i \hbar \ptl_{t_1} U_r(t_1, t_2) = \ham_r(t_1) U_r(t_1, t_2)$, then it is related to that in the 
laboratory frame by
\beq
\label{eq:U4}
U(t_1, t_2) = V(t_1) U_r(t_1, t_2) V^{\dagger}(t_2).
\eeq
The connection between the two formulations is made if we choose the time dependent 
unitary transformation to be
\beq
\label{eq:V1}
V(t) = \exp[-\frac{i}{\hbar} \int_0^t d \tau \ham_0 (\tau)],
\eeq
such that the $\ham_0(t)$ term is ``gauged out'' in the rotating frame. In this case
$\ham_r(t)$ coincides with $\ham_p(t)$ given by Eq.~\eqref{eq:H2}, $U_r(t, 0)$ with 
$W(t, 0)$, and Eqs.~\eqref{eq:U2} and \eqref{eq:U4} become identical with $t_2=0$.

However, note that the first formulation is more versatile in the sense that it can be still
used when $\ham_1$ is the zeroth order term and $\ham_0$ is perturbative. In this case,
we simply exchange $\ham_0(t) \leftrightarrow \ham_1$ in 
Eqs.~\eqref{eq:U3} and~\eqref{eq:H2}. The resulting expansion will not match with that in the 
rotating frame.

\subsection{Floquet perturbation theory}
\label{subsec:fpt}
Until now the discussion has been general, and it applies to all time dependent problems.
In the particular case of a Floquet system, where the time dependence is due a periodic 
external drive, we are interested in the stroboscopic time evolution operator 
$U(T, 0)$, where $T$ is the period of the drive. The related Floquet Hamiltonian
is defined by
\beq
\label{eq:Hf}
\ham_F \equiv \frac{i \hbar}{T} \log U(T, 0) =  \frac{i \hbar}{T} \log [ U_0(T, 0) W(T, 0)].
\eeq
We suppose that there is a small parameter that justifies the expansion
$W(T, ) = 1 + W_1(T) + W_2(T) + \cdots$, and correspondingly 
$\ham_F = \ham_F^{(0)} +   \ham_F^{(1)} +  \ham_F^{(2)} + \cdots$. Then, using 
Eq.~\eqref{eq:W2} and after some algebra the first few terms in the expansion of the 
Floquet Hamiltonian are given by
\begin{align}
 \ham_F^{(0)} &= \frac{i \hbar}{T} \log U_0(T, 0),
 \label{eq:HF0}\\
  \ham_F^{(1)} &= \frac{i \hbar}{T} W_1(T) = \frac{1}{T} \int_0^T d \tau \ham_p(\tau),
  \label{eq:HF1}\\
   \ham_F^{(2)} &= \frac{i \hbar}{T} \left[ W_2(T) - \frac{1}{2} W_1(T)^2 \right]
   \nonumber\\
   &= \frac{-i}{2 \hbar T} \int_0^T d \tau_1 \int_0^{\tau_1} d \tau_2 
   \left[ \ham_p(\tau_1) , \ham_p(\tau_2) \right],
   \label{eq:HF2}\\
   \ham_F^{(3)} &=  \frac{i \hbar}{T} \left[ W_3(T) - \frac{1}{2} \left( W_1(T) W_2(T)
   + W_2(T) W_1(T) \right) \right. \nonumber\\
   &+ \left. \frac{1}{3} W_1(T)^3 \right]
   \nonumber\\
   &= - \frac{1}{6 \hbar^2T} \int_0^T d \tau_1 \int_0^{\tau_1} d \tau_2 \int_0^{\tau_2} d \tau_3
   \left\{ \left[ \ham_p(\tau_1) , 
   \right.  \right. \nonumber\\
   &\left. \left. \left[ \ham_p(\tau_2) , \ham_p(\tau_3) \right] \right] 
   + \left[ \left[\ham_p(\tau_1) ,  \ham_p(\tau_2)  \right], \ham_p(\tau_3)  \right] 
   \right\}.
   \label{eq:HF3}
\end{align}
Eqs.\ \ref{eq:HF0}-\ref{eq:HF3} indicate the nested commutator structure of the higher-order terms 
of the Floquet Hamiltonian; we shall use them for explicit computation of $H_F$ in Sec.\ \ref{flo}. 

\section{Computation of the Floquet Hamiltonian}
\label{flo}

In this section we first provide analytical results for higher order terms in the Floquet
Hamiltonian for a cosine drive protocol in Sec.\ \ref{subsec:cosine}. This is followed, 
in Sec.\ \ref{genpro}, by a derivation and analysis of the first order Floquet Hamiltonian $H_F^{(1)}$ for a more 
general drive protocol.  

\subsection{Cosine modulation of interaction}
\label{subsec:cosine}

Consider a driven system described by Eq.~\eqref{eq:H1} where
\begin{align}
\ham_0(t) &= V_1 \cos \omega_D t \sum_i \hat{n}_i \hat{n}_{i+1}, 
\label{eq:H4}\\
\ham_1 &= \sum_i \left[ -J (\cda_i c_{i+1} + \rm{h.c.}) + V_0 \hat{n}_i \hat{n}_{i+1}
+ V_2 \hat{n}_i \hat{n}_{i+2} \right],
\label{eq:H5}
\end{align}
with $V_1 \gg (J, V_0, V_2)$. Thus, in the following we treat the $\ham_0$ term exactly, and 
$\ham_1$ perturbatively. 

Following Sec.~\ref{sec:formalism}, we have, 
using Eq.~\eqref{eq:U3},
\beq
\label{eq:U5}
U_0(t, 0) = \exp[-i \lambda \hat{B} \sin \omega_D t ],
\eeq
where $\lambda \equiv V_1/(\hbar \omega_D)$ is a dimensionless parameter and 
$\hat{B} \equiv \sum_j \hat{n}_j \hat{n}_{j+1}$.

The next step is to compute $\ham_p(t)$ using Eq.~\eqref{eq:H2}. As an intermediate step
we find
\begin{align}
\left[ \hat{B}, \ham_1 \right] &= -J \sum_i \hat{A}_i \left( \cda_i c_{i+1} - \cda_{i+1} c_i \right),
\label{eq:B1}\\
\left[ \hat{B} , \left[ \hat{B}, \ham_1 \right] \right] &= 
 -J \sum_i \hat{A}_i^2 \left( \cda_i c_{i+1} + \cda_{i+1} c_i \right),
 \label{eq:B2}
 \end{align}
 and so on, where
 \beq
 \label{eq:A}
 \hat{A}_i = \hat{n}_{i-1} - \hat{n}_{i+2}.
 \eeq.
 Using these relations we obtain
 \begin{align}
 \label{eq:H6}
 \ham_p(t) &= \exp[i \lambda \hat{B} \sin \omega_D t ] \ham_1 \exp[-i \lambda \hat{B} \sin \omega_D t ]
 \nonumber\\
 &= \ham_1 + i \lambda \sin \omega_D t \left[ \hat{B}, \ham_1 \right] 
 + \frac{1}{2!} ( i \lambda \sin \omega_D t )^2 
 \nonumber \\
 &\times \left[ \hat{B},   \left[ \hat{B}, \ham_1 \right] \right]
 + \frac{1}{3!} ( i \lambda \sin \omega_D t )^3 \left[ \hat{B} , 
 \left[ \hat{B},   \left[ \hat{B}, \ham_1 \right] \right] \right] 
 \nonumber \\
 &+ \cdots
 \nonumber\\
 &= \sum_i \left[ -J \left(e^{i \lambda \hat{A}_i \sin \omega_D t } \cda_i c_{i+1} + 
 e^{- i \lambda \hat{A}_i \sin \omega_D t}  \cda_{i+1} c_i  \right) \right.
  \nonumber \\
  &\left.   + 
  V_0 \hat{n}_i \hat{n}_{i+1}  +  V_2 \hat{n}_i \hat{n}_{i+2} \right].
 \end{align}
 Using the explicit form of $\ham_p(t)$ it is possible to compute order by order the 
 Floquet Hamiltonian.
 
 The zeroth order Floquet Hamiltonian $\ham_F^{(0)}$ vanishes because, from 
 Eq.~\eqref{eq:U5}, we have $U_0(T, 0) = 1$. 
 
 To compute the first order Floquet Hamiltonian we use the relation
 \begin{align}
 I_1(\hat{A}, \lambda) &\equiv \frac{1}{T} \int_0^T d \tau e^{i \lambda \hat{A} \sin (\omega_D \tau)}
 \nonumber\\
 &= J_0(\lambda \hat{A}) = (1 - \hat{A}^2) + \hat{A}^2 J_0 (\lambda),
 \end{align}
 where $J_n(x)$ is a Bessel function of the first kind with integer order. Using this relation and 
 Eq.~\eqref{eq:HF1} we get
 \begin{align}
 \label{eq:HF1-2}
 \ham_F^{(1)} &= \sum_i \left[ -J J_0(\lambda \hat{A}_i)  \left( \cda_i c_{i+1} + {\rm h.c.} \right)
 \right. \nonumber\\
 &+ \left. V_0 \hat{n}_i \hat{n}_{i+1} + V_2 \hat{n}_i \hat{n}_{i+2} \right].
 \end{align}
 If the drive frequency is tuned to $\omega_m$ such that 
 $\lambda_m = V_1/(\hbar \omega_m)$ coincides with the $m^{{\rm th}}$ zero of the Bessel
 function $J_0$, then the corresponding first order Floquet Hamiltonian is
 \begin{align}
 \label{eq:HF1-3}
 \ham_F^{(1)}(\lambda =\lambda_m) 
 &= \sum_i \left[ -J (1 - \hat{A}_i^2) \left( \cda_i c_{i+1} + {\rm h.c.} \right)
 \right. \nonumber\\
 &+ \left. V_0 \hat{n}_i \hat{n}_{i+1} + V_2 \hat{n}_i \hat{n}_{i+2} \right].
 \end{align}
 The above defines a model with constrained hopping, where only those hops are allowed
 which preserve the total number of nearest neighbors 
 $\hat{N}_D \equiv \sum_i \hat{n}_i \hat{n}_{i+1}$. This model is known to show 
 strong Hilbert space fragmentation \cite{fraglit2}.

The second order Floquet Hamiltonian can be broken into two parts
$\ham_F^{(2)} =  \ham_F^{(2a)} + \ham_F^{(2b)}$, with
\beq
\label{eq:HF-2a}
\ham_F^{(2a)} = \frac{-i}{2 \hbar T} \int_0^T d \tau_1 \int_0^{\tau_1} d \tau_2
\left[ \tilde{\ham}_p (\tau_1) \, , \, \tilde{\ham}_p (\tau_2)  \right],
\eeq
and
\begin{align}
\label{eq:HF-2b}
\ham_F^{(2b)} &= \frac{-i}{2 \hbar T} \int_0^T d \tau_1 \int_0^{\tau_1} d \tau_2
\left\{ \left[ \tilde{\ham}_p (\tau_1) \, , \, \hat{K} \right] \right.
\nonumber\\
&+ \left.  \left[ \hat{K} \, , \, \tilde{\ham}_p (\tau_2)  \right] \right\}.
\end{align}
In the above
\beq
\tilde{\ham}_p (\tau) \equiv -J \sum_i \left( e^{i \lambda \hat{A}_i \sin \omega_D \tau}
\cda_i c_{i+1} + {\rm h.c.} \right),
\eeq
and
\beq
\hat{K} = \sum_i \left( V_0 \hat{n}_i \hat{n}_{i+1} + V_2 \hat{n}_i \hat{n}_{i+2} \right).
\eeq
The details of the evaluation of the two parts is given in the appendix. The final result is
\beq
\label{eq:HF-2-2}
\ham_F^{(2)} = \frac{2 J \mathcal{C}(\lambda)}{\hbar \omega_D} \left[ 
\sum_i \hat{A}_i \left( \cda_i c_{i+1} - {\rm h.c.} \right) \, , \, \ham_F^{(1)} \right],
\eeq
where
\[
\mathcal{C}(\lambda) \equiv \sum_{k = 0}^{\infty} \frac{J_{2k +1}(\lambda)}{2k +1}.
\]

This concludes our derivation of the Floquet Hamiltonian for the cosine protocol. We note that 
${\mathcal H}_F^{(2)}$ does not respect the constrained hopping structure of ${\mathcal H}_F^{(1)}$ and therefore 
destroys HSF in the driven model beyond a prethermal timescale; below this timescale 
${\mathcal H}_F^{(1)}$ dominates the dynamics leading to prethermal relaization of HSF.

\subsection{An experimentally relevant drive protocol}
\label{genpro}

A possible realization of a standard fermion chain where coherent
quantum dynamics can be studied involves ultracold atom platforms
\cite{coldatom1,coldatom2}. In such realizations, both the hopping
amplitude and the nearest-neighbor interaction between the fermions
depend on the strength of the external lasers; therefore it is
difficult to dynamically alter one keeping the other fixed.
Therefore an experimental realization of strong HSF would require a
protocol which allows for simultaneous variation of both the hopping
and the interaction strength.

To take such simultaneous variations into account, we now consider a
fermionic chain with the Hamiltonian
\begin{eqnarray}
H &=& -J(t) \sum_{j} \left(c_j^{\dagger} c_{j+1} + {\rm h.c.}
\right) + (V_0 + V(t)) \sum_j \hat n_j \hat n_{j+1} \nonumber\\
&& + V_2 \sum_{j} \hat n_j \hat n_{j+2} \label{ham1}
\end{eqnarray}
where $J(t)$ and $V(t)$ are amplitudes of nearest neighbor hopping
and interactions respectively, $V_2 \ll |V(t)|$ is the amplitude of
the second-neighbor interactions, $c_j$ denotes the fermion
annihiliation operator on the $j^{\rm th}$ site of the chain, and
$\hat n_j = c_j^{\dagger} c_j$ is the fermion density operator.

In what follows, we choose a square pulse protocol so that
\begin{eqnarray}
V(t) &=& - V_1 \quad  t\le T/2, \nonumber\\
&=& V_1 \quad  T/2 < t \le T   \label{vprot} \\
J(t) &=& J_1  \quad  t\le T/2, \nonumber\\
&=& J_2 \quad  T/2 < t \le T,   \label{jprot}
\end{eqnarray}
with $V_1 \gg J_1, J_2, V_0, V_2$ so that one can reliably apply
FPT to compute the Floquet Hamiltonian. We note that the protocol
given by Eqs.\ \ref{vprot} and \ref{jprot} allows for simultaneous
variation of the hopping and the interaction strengths of the
fermions.

To obtain an analytic expression for the first-order Floquet Hamiltonian, we
first write the Hamitonian given by Eq.\ \ref{ham1} as $H= H_0 +
H_1$ where $H_0= V(t) \sum_j \hat n_j \hat n_{j+1}$ and
\begin{eqnarray}
H_1 &=& -J(t) \sum_{j} \left(c_j^{\dagger} c_{j+1} + {\rm h.c.}
\right) + V_0 \sum_j \hat n_j \hat n_{j+1}\nonumber\\ &&+ V_2 \sum_{j} \hat n_j \hat n_{j+2}. \label{pertham}
\end{eqnarray}
We then follow the standard procedure and obtain the evolution operator
corresponding to the term $H_0$ \cite{fptrev,fpt1,fpt2}. This yields
\begin{eqnarray}
U_0(t,0) &=&  e^{i V_1 t \sum_j \hat n_j \hat n_{j+1}/\hbar}
\quad\quad\quad t \le T/2  \label{zerofl1} \\
&=& e^{i V_1(T-t) \sum_j \hat n_j \hat n_{j+1}/\hbar}
\quad T/2 < t  \le T \nonumber
\end{eqnarray}
The Floquet Hamiltonian corresponding to $U_0 (T,0)$ can be easily read
off from Eq.\ \ref{zerofl1} to be identically $H_F^{(0)}=0$.

Next, we consider the effect of the terms in $H_1$ using
perturbation theory. The first order contribution to the evolution
operator from $H_1$ is given by
\begin{eqnarray}
U_1(T,0) &=&  \frac{-i}{\hbar} \int_0^T dt \, U_0^{\dagger}(t,0) H_1
U_0(t,0) \label{u1eq1}
\end{eqnarray}
To obtain analytic expression of $U_1(T,0)$ we first note that the
interaction terms in $H_1$ (Eq.\ \ref{pertham}) commute with $U_0$.
Thus the contribution of this term to $U_1$ is trivially obtained
and yields
\begin{eqnarray}
U_{1a}(T,0) &=& \frac{-i T}{\hbar} \left( V_0 \sum_j \hat n_j \hat n_{j+1} + V_2 \sum_j \hat n_j \hat n_{j+2}\right) \nonumber\\
\end{eqnarray}
In contrast, the contribution from the hopping term in $H_1$
requires a more detailed analysis. To this end, using Eq.\
\ref{zerofl1}, we write
\begin{widetext}
\begin{eqnarray}
&& U_{1b}(T,0) = \frac{i J_1}{\hbar} \int_0^{T/2} e^{-i V_1 t
\sum_j \hat n_j \hat n_{j+1}/\hbar} \sum_j (c_j^{\dagger} c_{j+1}
+{\rm h.c.}) e^{i V_1 t \sum_j \hat n_j \hat n_{j+1}/\hbar}
\nonumber\\
&& + \frac{i J_2}{\hbar} \int_{T/2}^T e^{-i V_1 T \sum_j \hat n_j
\hat n_{j+1}/\hbar} e^{i V_1 t \sum_j \hat n_j \hat
n_{j+1}/\hbar} \sum_j (c_j^{\dagger} c_{j+1} +{\rm h.c.}) e^{-i V_1 t \sum_j \hat n_j \hat n_{j+1}/\hbar} e^{i V_1 T \sum_j \hat
n_j \hat n_{j+1}/\hbar},  \label{u1eq2}
\end{eqnarray}
\end{widetext}
where we have used Eqs.\ \ref{vprot} and \ref{jprot}.

To evaluate Eq.\ \ref{u1eq2}, we note that the hopping from site $j$
to $j+1$ costs an energy due to the nearest-neighbor interaction if
it changes the number of bonds on the lattice whose both ends have
sites occupied by fermions. This allows us to define an operator
\begin{eqnarray}
\hat A_j  &=& \hat n_{j+2} - \hat n_{j-1}  \label{aop}
\end{eqnarray}
which takes values $\pm 1$ or $0$ on any site. The hopping of a
fermion from a site $j$ changes the energy due to nearest-neighbor
interaction only if $\hat A_j \ne 0$. This allows us to write
\begin{widetext}
\begin{eqnarray}
&& U_{1b}(T,0) = \frac{i J_1}{\hbar} \int_0^{T/2} dt \sum_j (e^{-i
V_1 t \hat A_j /\hbar} c_j^{\dagger} c_{j+1} +{\rm h.c.}) +
\frac{i J_2}{\hbar} \int_{T/2}^T dt \sum_j (e^{-i V_1 (T -t)
\hat A_j/\hbar} c_j^{\dagger} c_{j+1} +{\rm h.c.})\label{u1eq3}
\end{eqnarray}
\end{widetext}
Carrying out the integrals in Eq.\ \ref{u1eq3} and noting that $A_j$
can take values $0$ and $\pm 1$, we find \cite{ourprl}
\begin{widetext}
\begin{eqnarray}
U_{1b}(T,0) &=& \frac{iT}{\hbar} J_s \sum_j \left( \left[(1-\hat A_j^2) +
\hat A_j^2 e^{-iV_1 \hat{A}_j T/(4 \hbar)} \frac{ \sin V_1
T/(4\hbar)}{V_1 T/(4\hbar)} \right] c_j^{\dagger}
c_{j+1} + {\rm h.c.} \right) = \frac{-i T \hat{J}_c}{\hbar} \label{u1eq4}
\end{eqnarray}
\end{widetext}
where $J_s= (J_1+ J_2)/2$ and the
expression of $\hat{J}_c$ can be read off from Eq.\ \ref{u1eq4}. Thus we
find that for
\begin{eqnarray}
V_1 &=& 2 m \hbar \omega_D,  \label{cond1}
\end{eqnarray}
where $m \in Z$, the first order contribution to $U_1$ occurs
only if $\hat A_j =0$. This in turn means that the first order
evolution operator receives finite contribution from a constrained
hopping term which propagates fermion hopping in such systems. This
leads to a Floquet Hamitlonian that exhibits Hilbert space
fragmentation similar to that derived in Ref.\ \onlinecite{ourprl}.

\begin{figure}
\begin{center}
\includegraphics[width=0.98\linewidth]{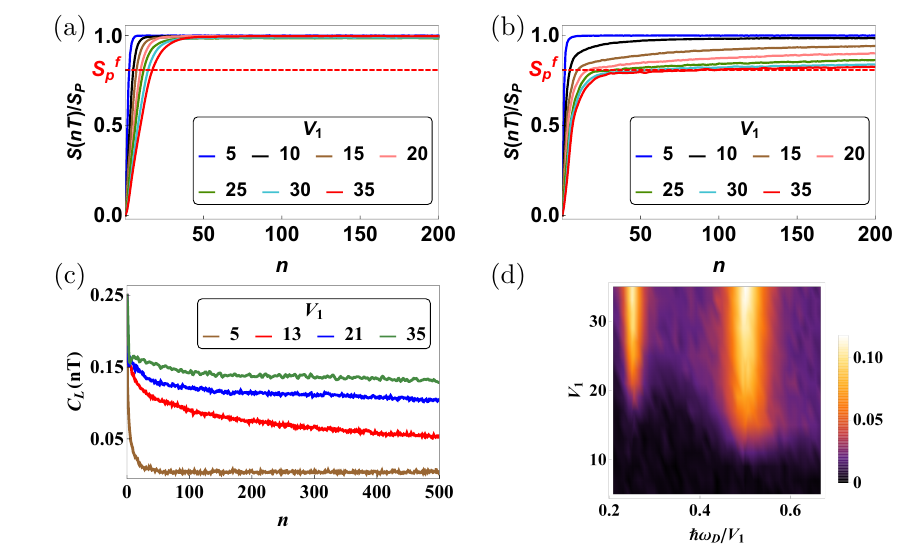}
\caption{(Color Online) (a) Plot of $S(nT)/S_p$ as a function of $n$ at $\omega_D=V_1/\hbar$ starting from a random Fock state for different values of the drive amplitude $V_1$. For all values of $V_1$, $S(nT)$ saturates to $S_p$. (b) Similar to (a) but at the special frequency $\omega_D=V_1/2\hbar$. With increase in $V_1$, $S(nT)$ saturates to $S_p^f$, the Page value of the fragment of $H_F^{(1)}$ from which the initial Fock state is chosen, for $n\leq 200$. (c) Plot of the density-density autocorrelator $C_L(nT)$ as a function of $n$ at $\omega_D=V_1/2\hbar$ starting from a infinite temperature thermal state. In this case too, the autocorrelator does not reach its thermal value of zero within the first 500 drive cycles, thus bearing signatures of prethermal HSF. (d) Value of $C_L(nT)$ after $n=5000$ drive cycles as a function of $V_1$ and $\hbar\omega_D/V_1$. The plot shows two special frequencies at $\hbar\omega_D/V_1=0.25$ and $\hbar\omega_D/V_1=0.5$. The time evolutions are performed using the exact unitary evolution operators, corresponding to drive protocols (\ref{vprot}) and (\ref{jprot}). The system sizes are $L=16$ for plots (a) and (b) and $L=14$ for (c) and (d). For all plots, $J_1=J_2/3= 0.5$ and $V_0=V_2=2$.} \label{fig0}
\end{center}
\end{figure}

The derivation of the first order Floquet Hamiltonian from Eq.\
\ref{u1eq4} and \ref{zerofl1} can be carried out in a straightforward manner \cite{fptrev}
and yields
\begin{eqnarray}
H_F^{(1)} &=&  \hat{J}_c +V_0 \sum_j \hat n_j \hat n_{j+1}+V_2 \sum_j \hat n_j \hat n_{j+2}
\label{hf1eq}
\end{eqnarray}
Thus the fragmentation exhibited by $H_F^{(1)}$ for this protocol is identical to that found in Ref.\
\onlinecite{ourprl}. 
In addition, it also allows for variation of $J$ which makes the
protocol much less restrictive compared to its counterpart in Ref.\
\onlinecite{ourprl}.

The corresponding dynamical signatures in the half-chain entanglement entropy and the density-density autocorrelation function are shown in Fig.\ \ref{fig0}. For these plots, we use Eqs.\ \ref{vprot} and \ref{jprot} and set the hopping amplitudes to $J_1=0.5$ for the first half of the drive and $J_2=1.5$ for the next half cycle in Fig.\ \ref{fig0}. Also, we set $V_0=V_2=2$. Figs.\ \ref{fig0}(a) and \ref{fig0}(b) show the evolution of the half-chain entanglement entropy starting from a random Fock state from the half-filled sector in a chain of length $L=16$ with periodic boundary condition at $\omega_D=V_1/\hbar$ (generic frequency) and $\omega_D=V_1/2\hbar$ (special frequency satisfying the relation in Eq. \ref{cond1}) respectively. Fig.\ \ref{fig0}(a) shows that away from the special frequency, for all values of the drive amplitude, the entanglement entropy $S(nT)$ saturates to the Page value $S_p$ of the half-filled symmetry sector from which the initial state is chosen, as is expected of ergodic systems. In contrast to this, Fig.\ \ref{fig0}(b) shows that at the special frequency, the entanglement entropy fails to reach $S_p$ with increasing drive amplitude within the first $200$ drive cycles. Instead, with increase in drive amplitude, $S(nT)$ saturates to $S_p^f$, the Page value of the fragment of $H_F^{(1)}$, from which the initial state is chosen. Both $S_p$ and $S_p^f$ have been computed analytically and numerically following Ref. \onlinecite{ourprl}.

Fig.\ \ref{fig0}(c) shows similar behavior for the time evolution of the density-density autocorrelator
\begin{eqnarray}
    C_L(nT)=\langle (n_L(nT)-1/2)(n_L(0)-1/2)\rangle
    \label{autocorr}
\end{eqnarray}
in an infinite temperature thermal state for a chain of length $L=14$ with open boundary condition at the special frequency $\omega_D=V_1/2\hbar$. A careful look at Eq. \ref{autocorr} reveals that $C_L$ also represents the connected autocorrelator since $\langle n_L(0)-1/2\rangle = 0$. Thus, in an ergodic system, $C_L(nT)$ is expected to decay to zero at long enough times signifying loss of any initial memory. However, Fig.\ \ref{fig0}(c) shows that with increasing drive amplitude $V_1$, $C_L$ saturates to a value much higher than zero at long enough times. This can be explained by considering the fragmented structure of $H_F^{(1)}$ and the Mazur's bound on the autocorrelator in the presence of the fragmented structure \cite{ourprl}. In \cite{ourprl}, we had seen that the long-time saturation value of the autocorrelator was above the lower bound predicted by the Mazur's bound. The autocorrelator decays down to zero when the chain is driven away from the special frequencies. Fig.\ \ref{fig0}(d) elucidates this by plotting the value of $C_L(nT)$ after $5000$ drive cycles as a function of the drive amplitude and the drive frequency. This plot can also serve as a ``phase diagram" in the drive frequency and drive amplitude space, where non-zero saturation values of $C_L(nT)$ (bright regions in the color plot) indicate parameter regimes where prethermal fragmentation is observed.

It is to be noted here that for a given drive amplitude, the rate of thermalization is faster as compared to that reported in [\onlinecite{ourprl}]. This is to be attributed to the asymmetric drive protocol (different values of the hopping amplitude during the two half-cycles) used here. Due to the asymmetric nature of the protocol, the lowest non-trivial correction to the constrained Hamiltonian at the special frequency comes from the second-order Floquet Hamiltonian, $H_F^{(2)}$ as compared to $H_F^{(3)}$ in [\onlinecite{ourprl}]. This, in turn, results in a shorter thermalization timescale.

\section{Other filling fractions}
\label{ofil}

\begin{figure}
\begin{center}
\includegraphics[width=0.98\linewidth]{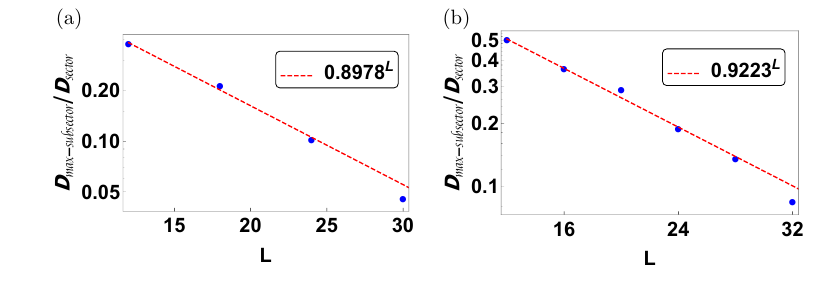}
\caption{(Color Online) (a) Plot of the ${\mathcal D}_{\rm
sub-sector}/{\mathcal D}_{\rm sector}$ for $H_F^{(1)}$ as a function
of $L$ for $N/L=1/3$ showing exponential reduction with $L$ similar
to the half-filling case. (b) Similar plot for $N/L=1/4$. For both
plots, $J=1$ and $V_1/(\hbar \omega_D)=2m$.} \label{fig1}
\end{center}
\end{figure}

In this section, we study the driven fermion chain away from
half-filling to demonstrate the robustness of the fragmentation
signature. To this end, we consider the driven fermion chain at
filling fractions $N/L=1/3, 1/4$ (where $N$ is the fermion number
and $L$ is the chain length). In what follows, we shall use the 
square-pulse protocol given by $V(t) = -(+) V_1$ for $t\le(>) T/2$ in accordance 
with Ref.\ \onlinecite{ourprl}.

We begin our study by analyzing the Hilbert space dimension (HSD) of
the largest fragment of the first order Floquet Hamiltonian (Eq.\ \ref{eq:HF1}) 
at $N/L=1/3$ and $1/4$. This is shown in Fig.\
\ref{fig1} where the ratio of the HSD of the largest fragment,
${\mathcal D}_{{\rm sub-sector}}$, and the total HSD of the symmetry
sector (one-third filled sector in (a) and one-fourth filled sector in (b)), ${\mathcal D}_{{\rm sector}}$, is plotted as a function of
$L$ for $J=1$ and at a special frequency $V_1/(\hbar \omega_D)= 2m$, where $m$ is an integer. We find a clear
signature of exponential decay of this ratio for both $1/3$ and
$1/4$ filling fractions as a function of $L$. This indicates the
possibility of the presence of signature of strong HSF in the
dynamics of the driven chain at these filling.

\begin{figure}
\begin{center}
\includegraphics[width=0.98\linewidth]{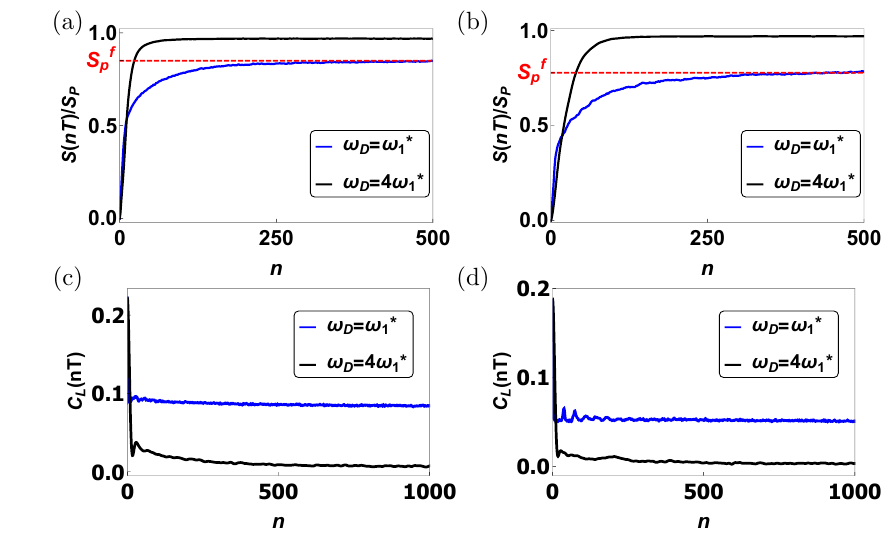}
\caption{(Color Online) (a) Growth of entanglement entropy from
exact dynamics for $L=18$ and $N = 6$, starting from a randomly
chosen Fock state. The result is averaged over 10 such states chosen
from the same fragment of the first order Floquet Hamiltonian. The
sector dimension from which the state is chosen is $1980$. The
entanglement entropy is scaled by the Page value $S_p$ of the
symmetry sector for which $N/L=1/3$. At the special frequencies, EE
saturates to a value less than $S_p$ but close to $S_p^f$ for the
fragment from which the initial state was chosen. (b) Same as in (a)
but for $L=20$ and $N=5$ corresponding to $1/4$ filling. The sector
dimension of the fragment from  which the initial state is chosen is
$1050$. (c) Plot of $C_L(nT)$ as a function of $n$ at the special
frequency $V_1/(\hbar \omega_d)=2$ and away from it $V_1/(\hbar
\omega_D)=1/2$ for $N/L=1/3$ and $L=18$ The initial state is same as
in (a). (d) Same as (c) but for $L=20$ and $N/L=1/4$; the initial
state is same as in (b). For all plots $J=1$.} \label{fig2}
\end{center}
\end{figure}

To verify this expectation, we compute the entanglement entropy
$S(nT)$ as a function of $n$ and at different drive frequencies.
For an ergodic driven system, $S$ is expected to increase with $n$
and eventually saturate to its Page value corresponding to the
symmetry sector $S_p$ 
irrespective of the initial state chosen for the
dynamics\cite{page1}. In contrast, for a chain which exhibits HSF,
$S(nT)$ is expected to saturate to the Page value of the fragment to
which the initial state belongs: $S \to S_{p}^f$. 
Thus the saturation value of $S$ is
lower; also it depends on the initial state from which the dynamics
originates. This allows one to distinguish between dynamical
behavior of a driven chain with and without strong HSF. A plot of $S(nT)/S_p$, 
shown in Fig.\ \ref{fig2}(a) for $N/L=1/3$
and Fig.\ \ref{fig2}(b) for $N/L=1/4$, clearly shows the distinction
between the behavior of $S$ at and away from the special
frequencies. $S(nT)/S_p$ saturates, for both fillings, to unity at
large $n$ away from the special frequencies($V_1/(\hbar
\omega_D)=1/2$); in contrast, at the special frequency $V_1/(\hbar
\omega_D)=2$, they saturate to a lower value which corresponds to
$S_p^f$ of the respective sectors from which the initial states are
chosen.

In addition, we compute the density-density autocorrelation function given by Eq.\ \ref{autocorr}.  
Fig.\ \ref{fig2}(c) and (d) show the behavior of $C_L(nT)$
as a function of $n$ at and away from the special frequency for
$N/L=1/3$ and $1/4$ respectively. We find that in both cases,
$C_L(nT)$ saturates to finite value at the special frequency and to
zero away from it. Thus these plots confirm the existence of strong
HSF similar to the half-filling sectors in these fermionic chains.

\begin{figure}
\begin{center}
\includegraphics[width=0.98\linewidth]{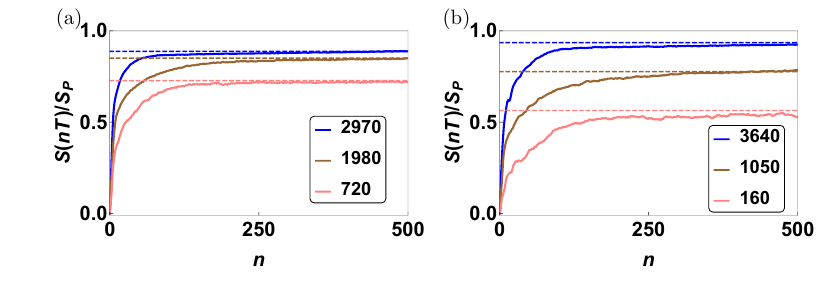}
\caption{(Color Online) Plot of $S(nT)/S_p$ as a function of $n$ at
$V_1/(\hbar \omega_D)=2$ starting from Fock states belonging to
different fragments for (a) $L=18$ and $N=6$ and (b) $L=20$ and
$N=5$. The numbers label the Hilbert space dimensions of the
fragments and the dashed lines indicate the corresponding $S_p^f$.
For all plots $J=1$.} \label{fig3}
\end{center}
\end{figure}

Finally, in Fig.\ \ref{fig3}, we show the plot of $S(nT)/S_p$
starting from several initial Fock states which belong to different
sectors with different values of $S_p^f$ for $V_1/(\hbar
\omega_D)=2$. We find that in each case, both at $N/L=1/3$ (Fig.\
\ref{fig3}(a)) and $1/4$ (Fig.\ \ref{fig3}(b)), $S(nT)/S_p <1$ for
large $n$; moreover, $S(nT) \to S_p^f$ corresponding to the fragment
of $H_F^{(1)}$ to which the initial state belongs. This clearly
demonstrates signature of HSF at these filling fractions.

\section{Density-density out-of-time ordered correlation function}
\label{sec:otoc1}

In this section we analyze the out-of-time ordered correlator (OTOC) for the driven chain. The basic definitions are 
outlined in Sec.\ \ref{prelim1}. This is followed by numerical study of OTOC for a
chain with periodic boundary condition (PBC) and starting from an infinite temperature 
thermal state in Sec.\ \ref{pbot}. Finally we study the behavior of OTOC for fermion chains 
with open boundary condition (OBC) and starting from the ${\mathbb Z}_2 = |0,1,0,1,\ldots\rangle $ Fock state 
in Sec.\ \ref{obot}.

\subsection{Preliminary}
\label{prelim1}
The study of OTOC serves as an important tool to diagnose the rate of propagation of local information in a quantum system \cite{otocref1,otocref2,otocexp,otocrev}. Ergodic systems are known to exhibit ballistic spread of local information accompanied by a diffusive front. In case of non-ergodic systems, the behavior of information propagation, as detected using OTOC, ranges from logarithmic growth in many-body localized systems \cite{otocmbl} to alternate scrambling and unscrambling in certain integrable systems \cite{sur1}. For fragmented systems, the scrambling of information is expected to be slow since the Hamiltonian does not connect states belonging to different fragments; however, its detailed features, in the presence of a periodic drive, have not been studied earlier.

To probe the rate of scrambling of information in our system, we study the temporal (in stroboscopic times) and spatial profile of the OTOC 
\begin{eqnarray}
F(r, nT) &=& \langle \Tilde{n}_i(nT)\Tilde{n}_j(0)\Tilde{n}_i(nT)\Tilde{n}_j(0) \rangle,
\end{eqnarray} 
where $\Tilde{n}_i=2 n_i -1$ with $n_i$ and $n_j$ being the number density operators at sites $i$ and $j$ respectively and $r = |i-j|$ measures the distance between these two sites. We take the average with respect to both an infinite-temperature thermal state and the $|{\mathbb Z}_2\rangle$ state. Since the operator $\Tilde{n}_i$ is hermitian and squares to identity, it can be shown that the function $F(r,nT)$ is related to the squared commutator $C(r,nT)$ as 
\begin{eqnarray}
C(r,nT) &=& \langle [\Tilde{n}_i (nT), \Tilde{n}_j]^\dagger [\Tilde{n}_i (nT), \Tilde{n}_j]\rangle \nonumber\\
&=& 2(1-F(r,nT)). 
\end{eqnarray}
Cast in this form, it can be argued that as the operator $\Tilde{n}_i$, initially localized at site $i$, spreads to the site $j$, the value of $C(r,nT)$ at this site gradually increases from zero and hence the OTOC, $F(r,nT)$ decreases from $1$.  A higher value of $C(r,nT)$ (i.e. lower value of $F(r,nT)$) at a given instant of time therefore indicates larger spread of the local operator ($\tilde n_i$ in the present case). 

\subsection{Infinite-temperature initial state}
\label{pbot} 

\begin{figure}
    \centering
    \includegraphics[width=0.98\linewidth]{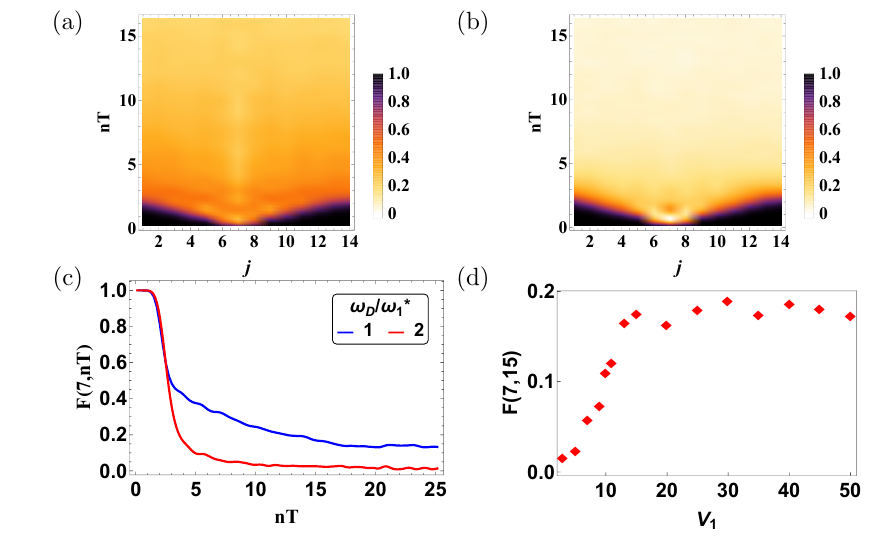}
    \caption{(Color Online) (a) Profile of the OTOC $F(r,nT)$ in an infinite temperature thermal state for $L=14$ (half-filled sector) with PBC and $\omega_D=V_1/\hbar$. The initial operator $\Tilde{n}_i$ is localized at the centre of the chain $i=7$. The index $j$ in the x-axis labels the site of the chain and $r=|j-7|$ for all plots. A ballistic spread of information is seen with the value of the OTOC rapidly dropping close to zero as is expected of a thermalizing system. (b) Same as (a) but at the special frequency $\omega_1^*=V_1/2\hbar$, where the first-order Floquet Hamiltonian $H_F^{(1)}$ is fragmented. Although there is some information scrambling in this case due to mixing of states within fragments, the value to which $F(r,nT)$ reaches at similar times is higher as compared to the thermalizing case. This implies that the extent of information scrambling is less in this case, compared to (a). (c) Profile of $F(r,nT)$ at site $j=14$ both at and away from the special frequency $\omega_1^*$, illustrating the same point. (d) Variation of the value of $F(r,nT)$ at the farthest site $j=14$ after $t=nT=15$ with drive amplitude $V_1$ and $\omega_D=\omega_1^{\ast}$. With decrease in $V_1$, higher-order terms in the pertubation series gain prominence and enhance information scrambling. For all plots, $J=1,V_0=V_2=2$. For (a)-(c), $V_1=50$. All the results are obtained using the exact time-evolution operator.}
    \label{fig4}
\end{figure}
In this section, we study the spread of OTOC in an infinite-temperature thermal state for a half-filled chain of length $L=14$ with PBC both at the special frequency ($\hbar \omega_1^*=V_1/2$, shown in Fig \ref{fig4}(b)) and away from it ($\omega_D=2\omega_1^*$, shown in Fig \ref{fig4}(a)). The operator is initially localized at the centre of the chain, i.e. $i=L/2$ in both the cases. Fig. \ref{fig4}(a) shows that at a generic frequency, the operator spreads ballistically. Such a spread can be inferred from the linear variation of $r$, for sites at which $F(r,t)$ has almost similar values, as a function of $t$. 

$F(r,nT)$ quickly falls to a value close to zero, implying that $C(r,nT)$ saturates to a value close to $1$. At the special frequency, however, Fig. \ref{fig4}(b) shows that although the local information reaches the farthest site almost at the same time as in the previous case, the OTOC saturates to a higher value as compared to its thermalizing counterpart. This is a direct consequence of the fact that the first order Floquet Hamiltonian $H_F^{(1)}$, which is fragmented, only allows mixing of the states within a particular fragment. Although the infinite-temperature thermal initial state (represented by a density matrix) weighs all the states equally, during time evolution they can only be connected with states belonging to the same fragment. As a result, they fail to spread out through the whole Hilbert space. The information is scrambled only due to mixing between states within individual  fragments; this leads to lower scrambling in the prethermal regime than that due to ergodic  evolution away from the special frequency. Fig. \ref{fig4}(c) illustrates this fact by plotting the value of $F(r,nT)$ for site $j=14$ both at and away from the special frequency. As the drive amplitude decreases, the higher order terms in the perturbation series start dominating, allowing mixing between different fragments. This enhances information scrambling leading to a decrease in the value of the OTOC. This is shown in Fig. \ref{fig4}(d) which plots the variation of the value of OTOC at the farthest site ($j=14$) at $nT=15$ as a function of $V_1$  for $\omega_D=\omega_1^{\ast}$. This shows that information scrambling is suppressed beyond a critical $V_1$ where signatures of fragmentation can be found over a long pre-thermal timescale. 


\subsection{$\mathbb{Z}_2$ State}
\label{obot} 
\begin{figure}
    \centering
    \includegraphics[width=0.98\linewidth]{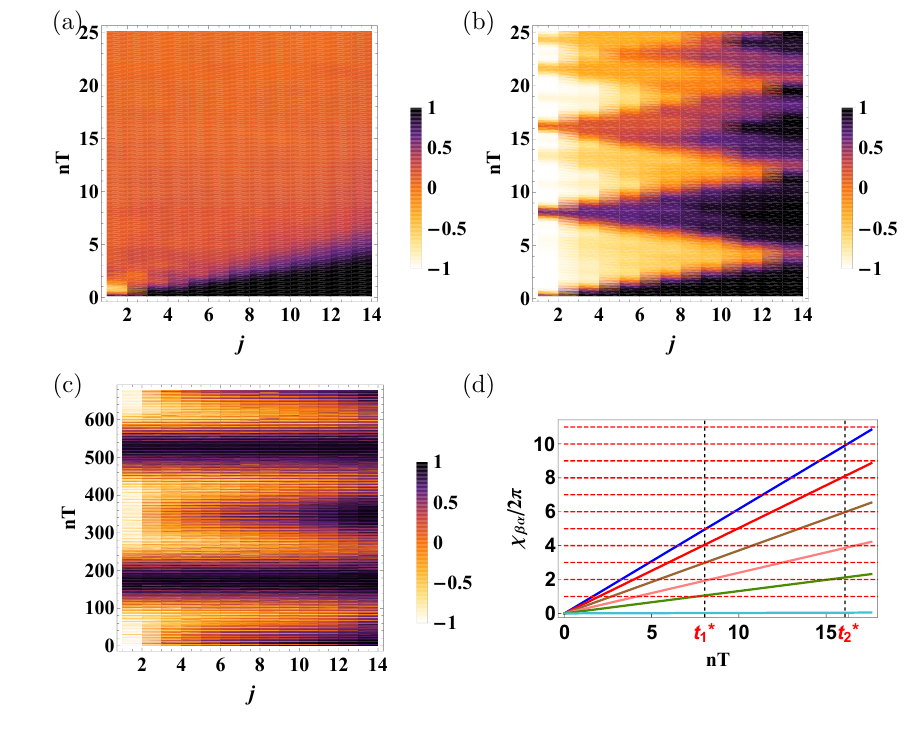}
    \caption{(Color Online) (a) Profile of the OTOC $F(r,nT)$ in a $\mathbb {Z}_2$ state for L=14 with OBC and at a generic frequency $\omega_D=V_1/\hbar$. The initial operator $\Tilde{n}_i$ is localized at one of the edges of the chain $i=1$. The information reaches to the other end of the system ballistically, bearing signature of thermalizing systems. (b) Same as (a) but at the special frequency $\omega_1^*=V_1/2\hbar$. The information after reaching the other end starts unscrambling again. This alternate scrambling and unscrambling of information continues over a short timescale, dictated by the quasienergy spectrum of the fragmented $H_F^{(1)}$ as explained in the main text. (c) Long-time oscillations in the profile of the OTOC at $\omega_1^*$. (d) Plot of $\chi_{\beta\alpha}(nT)$ (defined below Eq. \ref{otoc4} in the main text) as a function of $t=nT$. The color codes are: $\chi_{28}$ (blue), $\chi_{38}$ (red), $\chi_{48}$ (brown), $\chi_{58}$ (pink), $\chi_{68}$ (green) and $\chi_{78}$ (cyan). The red dashed lines mark integer multiples of $2\pi$. The first two times when these phases are very close integer multiples of $2\pi$, are marked as $t_1^*$ and $t_2^*$ respectively. These correspond to the first two recurrence times in (b). For all plots, $J=1, V_0=V_2=2, V_1=50$.}
    \label{fig5}
\end{figure}

In this section, we study the spatial and temporal profile of the OTOC in a $|{\mathbb Z}_2\rangle$ state at the special frequency with OBC. Fig. \ref{fig5}(a) shows that away from the special frequency, starting from one end of the chain, the information propagates ballistically to the other end, as is expected for a ergodic system; $F(r,t)$ monotonically decays to near-zero value at all sites within $nT\le 10$. In contrast, as shown in Fig. \ref{fig5}(b) and (c), at the special frequency
the behavior of $F(r, nT)$ is quite different and it shows signature of fragmentation. 
In Fig.\ \ref{fig5}(b), we find that at short time scales $nT \sim$ 10, there are initial 
fast oscillations which lead to alternate scrambling and unscrambling of information.
Such alternate scrambling and unscrambling of information is reminiscent of the 
behavior of OTOC in integrable systems \cite{sur1}; however, as we show below, 
the mechanism for this phenomenon is different in the present case.
Furthermore, over longer time scales $nT \sim$ 100 - 500, we find slow 
oscillatory behavior as seen in Fig.\ \ref{fig5}(c). As discussed below, this is related to 
tunneling between two near degenerate states.

Both the above oscillatory features can be related to the fact that
at high drive amplitude and at the special frequencies, the dynamics is mostly governed by $H_F^{(1)}$ at short and intermediate timescales. To understand the behavior of $F$, we therefore focus on the fragment of $H_F^{(1)}$ (with OBC) to which $|{\mathbb Z}_2\rangle$ belongs. For $L=14$, there are $8$ states in this fragment namely 
\begin{align}
\mathcal{H}=&\{|{\mathbb Z}_2\rangle, |j_h=2\rangle,  |j_h=4\rangle,  
|j_h=6\rangle,
\nonumber \\
&|j_h=8\rangle,  |j_h=10\rangle, |j_h=12\rangle, |\Bar{\mathbb Z}_2\rangle\}
\label{nd0states}
\end{align}
where $|\Bar{\mathbb Z}_2\rangle=|1,0,1,0,\ldots\rangle, \text{ and } |j_h\rangle$ is a state with one hole-defect (where aa hole-defect implies two adjacent unoccupied sites) at position $j_h$ and zero particle defect (i.e. no two adjacent sites are occupied), viz $|j_h= 2\rangle = |1,0,0,1,0,1,0,1,0,1,0,1,0,1\rangle$. 
Note, the constrained hopping introduces dynamics between these eight states, and 
$H_F^{(1)}$ in this subspace is equivalent to a nearest neigbor hopping model 
of a linear chain with eight sites and OBC. Here $|{\mathbb Z}_2\rangle$ and 
$|\Bar{\mathbb Z}_2\rangle$ form the ends of the chain while $j_h = 2, 4, \ldots, 12$ form
the sites in between.

The OTOC at a site $j$ will have the structure
\begin{equation}
    F(r_1,nT) = \langle \mathbb{Z}_2| \Tilde{n}_1(nT) \Tilde{n}_j(0) \Tilde{n}_1(nT) \Tilde{n}_j(0)|\mathbb{Z}_2\rangle
    \label{otoc1}
\end{equation}
where $r_1=|j-1|$. Inserting the complete set of states $|m\rangle$ from this fragment and noting that the operator $\tilde n$ is diagonal in the Fock basis, this expression reads
\begin{equation}
    F(r_1,nT)\approx\sum_m (-1)^j f^j_m |\langle m(nT)| \Tilde{n}_1 |\mathbb{Z}_2 (nT)\rangle|^2
    \label{otoc2}
\end{equation}
where $f^j_m = \langle m|\Tilde{n}_j|m\rangle$. Expanding $|\mathbb{Z}_2\rangle$ and $|m\rangle$ in the energy eigenstates of $H_F^{(1)}$: $|\mathbb{Z}_2\rangle = \sum_\alpha c_\alpha |\phi_\alpha\rangle$, $|m\rangle=\sum_\beta c_\beta^m |\phi_\beta\rangle$, Eq. \ref{otoc2} yields
\begin{equation}
    F(r_1,nT)\approx\sum_m (-1)^j f^j_m g_m (nT)
    \label{otoc3}
\end{equation}
where 
\begin{equation}
    g_m(nT)= \Big|\sum_{\alpha,\beta} c_\beta^{m*} c_\alpha e^{-i\chi_{\beta\alpha}(t)} N^1_{\beta\alpha}\Big|^2
    \label{otoc4}
\end{equation}
with $N^1_{\beta\alpha}=\langle\phi_\beta|\Tilde{n}_1|\phi_\alpha\rangle$ being the matrix element of $\Tilde{n}_1$ between the energy eigenstates and $\chi_{\beta\alpha}(t)=(\epsilon_\alpha-\epsilon_\beta)nT/\hbar$. 

\subsubsection{Short and Long time Oscillations}
In Eq. \ref{otoc3} the spatial dependence on $r_1$ or $j$ is factorized out from the time dependence $nT$. This implies that the time dependence of the OTOC is site-independent, i.e. the recurrence time at every site is the same and the recurrence happens when all the phases $\chi_{\beta\alpha}(t)$ are approximately close to integer multiples of $2\pi$. We arrange the spectrum $\epsilon_1<\epsilon_2<\ldots<\epsilon_8$. Numerically, we find that the matrix elements $N^1_{\beta\alpha}$ between the states $|\phi_\beta\rangle; \beta=2,3,\ldots,8$ and $|\phi_\alpha\rangle; \alpha=7,8$ are an order of magnitude higher than the rest of the off-diagonal matrix elements. 
This is because the states $\beta = 1, 2, \ldots, 6$ are mostly made of the six single-hole
wavefunction, while the states $\alpha = 7, 8$ are mostly made of the states
$\mathbb{Z}_2$ and $\bar{\mathbb{Z}}_2$. Since the last two states have one extra 
next-nearest-neighbor interaction compared to the first six, 
$\epsilon_{\alpha \beta} \sim V_2$, and this energy scale shows up in the fast oscillations 
seen over timescales $nT \sim$ 10.
Thus, 
in this relatively short time, the recurrence is 
predominantly dictated by the phases $\chi_{\beta\alpha}(t)$ with $\beta=2,3,\ldots,7$ and $\alpha=8$. Fig. \ref{fig5}(d) plots these phases as a function of $t=nT$. It can be seen that the recurrence occurs when all these phases are close to $2\pi p_0$ (where $p_0 \in Z$) as shown in Fig. \ref{fig5}(d). The first two of these times are marked with $t_1^*$ and $t_2^*$ in Fig.\ref{fig5}(d). These are not exactly periodic because of involvement of multiple phases in the dynamics. It is also to be noted from Fig. \ref{fig5}(d) that the energies $\epsilon_{7,8}$ (which are mostly linear combinations of 
$\mathbb{Z}_2$ and $\bar{\mathbb{Z}}_2$ Fock states) are so close that for the short timescale involved, the phase $\chi_{87}(t)$ almost remains close to zero; it does not play much role in determining the short recurrence time. 
Thus, from the above discussion it is clear that the recurrences at short 
timescales owe their existence to two features in the model. First, the finite 
next-nearest-neighbor interaction energy $V_2$. Second, the OBC which allows the 
single-hole states to be included within the same fragment as to which the states
$\mathbb{Z}_2$ and $\bar{\mathbb{Z}}_2$ belong (with PBC, the fragment has only the
$\mathbb{Z}_2$ and $\bar{\mathbb{Z}}_2$ states).

\begin{figure}
    \centering
    \includegraphics[width=1.0\linewidth]{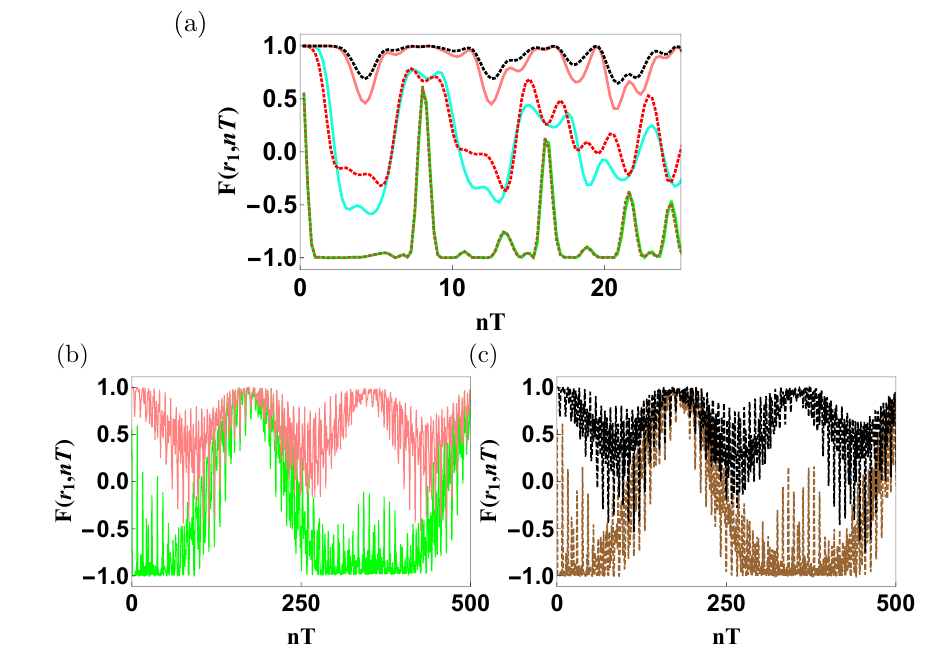}
    \caption{(Color Online) (a) Comparison of exact and approximate estimate (Eq. \ref{otoc3}) of $F(r_1,nT)$ for $j=1$ (green solid, brown dashed), $j=5$ (cyan solid, red dashed) and $j=14$ (pink solid, black dashed) sites. The solid lines represent results obtained from exact dynamics and dashed lines represent approximate estimates obtained using the fragment of $H_F^{(1)}$. The first two recurrence times are in good agreement, emphasizing the role of fragmentation in the scrambling and unscrambling behavior observed in Fig. \ref{fig5}(b). For $j=1$, the agreement between approximate and exact numerical result is nearly perfect and the green solid and the brown dashed lines are almost on top of each other. (b), (c) Similar comparison for the long time oscillations for $j=1, 14$, following the color scheme used in (a). (b) Plot of exact numerical result for $j=1, 14$. (c) Plot of approximate estimate for $j=1,14$.}
    \label{fig6}
\end{figure}

In Fig. \ref{fig6}(a), we show the comparison between results for $F(r_1,nT)$ obtained from exact dynamics (solid lines) and the analytical estimate obtained from $H_F^{(1)}$ in Eq. \ref{otoc3} (dashed lines) for some representative sites $j=1,5,14$. It can be seen that the first two recurrence times at $t_1^*=8.07, t_2^*=16.21$ are well approximated by Eq. \ref{otoc3}.

The phase $\chi_{87}(t)$ manifests itself only at longer time scales of $nT \sim$ 100 - 500. 
As seen in Fig.\ \ref{fig5}(c), over this 
time scale $F(r_1,nT)$ oscillates from values nearly one to nearly minus one with frequency 
$\Omega$, where $\Omega = (\epsilon_8 - \epsilon_7)/\hbar$. In Fig. \ref{fig6}(b) and (c) we show that 
these oscillations can be explained using Eq \ref{otoc3} by considering the $8$ states belonging to this fragment of $H_F^{(1)}$. In Appendix \ref{otocobclong}, we show that a four state ansatz can be used to arrive at this result for a high next-nearest-neighbor interaction strength, when the {$\mathbb Z_2$, $\Bar{\mathbb Z}_2$} states are well separated in energy from the remaining $|j_h\rangle$ states.

\subsubsection{Spatial profile of OTOC}
The spatial dependence in the profile of the OTOC appears through the term $h_{jm}=(-1)^j f^j_m$ in Eq. \ref{otoc3}. The initial linear increase in Fig. \ref{fig5}(b) for $nT\leq 5$ can be explained by focusing on this term. The profile of $h_{jm}$ for odd $j$'s reads
\begin{equation}
    h_{jm}=\begin{pmatrix}
        1 & -1 & -1 & -1 & -1 & -1 & -1 & -1\\
        1 & 1 & -1 & -1 & -1 & -1 & -1 & -1\\
        1 & 1 & 1 & -1 & -1 & -1 & -1 & -1\\
        1 & 1 & 1 & 1 & -1 & -1 & -1 & -1\\
        1 & 1 & 1 & 1 & 1 & -1 & -1 & -1\\
        1 & 1 & 1 & 1 & 1 & 1 & -1 & -1\\
        1 & 1 & 1 & 1 & 1 & 1 & 1 & -1\\
    \end{pmatrix}
    \label{nprofile}
\end{equation}
where the rows label the odd sites $j=1,3,5,\ldots 13$ and the columns label the different Fock states $|m\rangle$ in this fragment, in the same order as in Eq. \ref{nd0states}. The time dependent functions $g_m(nT)$ are positive definite at all times. As $j$ increases, the number of $g_m(nT)$'s having positive weights increases linearly as is evident from Eq. \ref{nprofile}. Thus, the shift of $F(r_1,nT)$ from $1$ happens progressively at a later time as $j$ increases. It is also useful to note that $f^{2k-1}_m = -f^{2k}_m$ for all $k \text{ and } m$, so that $h_{2k-1,m}=h_{2k,m}$ and hence at any given instant of time, $F(2k-1,nT)=F(2k-2,nT)$. Thus the even sites $j$ have similar behavior as the odd sites.

\section{Discussion}
\label{dissc}

In this work, we studied the dynamics of a periodically driven Fermi chain and extended the 
study of prethermal HSF in these system undertaken in Ref.\ \onlinecite{ourprl} in several ways. 

First, we have studied the existence of such prethermal HSF beyond half-filling in such chains. We found the existence of such 
prethermal HSF phase for several other filling fractions such as $N/L=1/4$ and $1/3$. This shows that the robustness of the pretheraml
MBL phase in such driven chain. 

Second, we provide a derivation of the first and second order Floquet Hamiltonian in such driven system in an alternative
manner. Our derivation brings out the commutator structure of the Floquet Hamiltonian; in particular, we find that the second order 
term in the Floquet Hamiltonian, $H_F^{2}$, can be expressed as a commutator of a constrained current operator $\sum_j A_j(c_j^{\dagger} c_j-h.c.)$ 
with $H_F^{(1)}$. We expect similar commutation relations to hold for higher order terms in $H_F$; this sheds light on the symmetry content of 
the higher order terms in the Floquet Hamiltonian for the cosine drive protocol. 

Third, we extend our analysis to experimentally relevant and slightly more complicated drive protocols. In a typical experiment, 
involving ultracold atoms, the interaction strength between fermions and their hopping strength are both controlled by intensities of the applied lasers. 
Consequently, experimentally relevant protocols must allow change of both the hopping amplitudes and interaction strength. We show that 
the prethermal HSF is stable for a large class of such drives and chart out a phase diagram for the special frequencies at which it occurs.

Finally, we study the behavior of density-density OTOC for such driven systems. Our study shows that such OTOCs can serve as detectors of 
such prethermal HSF in two distinct ways. First, irrespective of the boundary condition used, the OTOC $F(r,t)$ 
for a finite fermion chain driven at the special frequency and starting from a ${\mathbb Z}_2$ initial state, 
exhibits a larger long-time value than when driven away from such frequencies. In addition, it also exhibits oscillations with very large periodicity 
at the special frequencies that have the same origin as the correlation functions discussed in Ref.\ \onlinecite{ourprl}. In contrast, no such oscillations
are found when one is away from the special frequency; the OTOC monotonically decreases to zero. Second, for fermion chains with open boundary condition and 
driven at special frequencies, we find periodic scrambling and unscrambling of information which is in sharp contrast to standard behavior of OTOCs 
in driven ergodic systems. Such a behavior was found earlier for integrable spin chains \cite{sur1}; however, their origin for systems with prethermal HSF
quite different and can be tied to the localization of the driven system within a group of Fock states with same dipole number ($n_d=0$). For chains with open boundary condition, 
there are $O(L)$ such states which govern the dynamics up to a long prethermal time scale leading to periodic scrambling and unscrambling. This phenomenon is qualitatively 
different from the behavior of OTOC away from the special frequencies where it monotonically decays due to fast spread of the driven system through the Hilbert space; it is also different 
for a chain with PBC with two Fock states (${\mathbb Z}_2$ and ${\bar {\mathbb Z}_2}$) in the $n_d=0$ sector where no such unscrambling is found.  

Experimental verification of our result would require realization of isolated fermi chain. A possible scenario for this is a $1D$ fermion systems with nearest neighbor hopping
and local interaction realized suing ultracold fermions in an optical lattice. We propose to drive this with the experimentally relevant protocol discussed in this work; this 
can be achieved by varying strength of lasers used to generate the lattice \cite{coldatom1,coldatom2}. The simplest measurement would involve measuring $\langle \hat n_d\rangle$ 
as a function of time. We predict that the dynamics of $\langle n_d\rangle$ staring from the ${\mathbb Z}_2$ state for such a chain will be 
approximately constant (and close to zero) for a long prethermal timescale when the system is driven at the special frequency. This is to be contrasted the 
behavior of $\langle n_d\rangle$ away from the special frequency which should exhibit rapid dynamics at short timescale. 

In conclusion, we have studied several aspects of prethermal HSF in a driven Fermi chain. Our results have showed the robustness of this phenomenon 
by confirming its existence for different, experimentally relevant, drive protocols and also when the system is away from half-filling. In addition we have 
demonstrated that OTOCs may serve as a marker for such prethermal HSF; they exhibit periodic scrambling and unscrambling for fermion chains with open boundary 
condition driven at the special frequency.

{\it Acknowledgement}: SG acknowledges CSIR, India for support
through project 09/080(1133)/2019-EMR-I. IP thanks Edouard Boulat
for discussions. KS thanks DST, India for support through SERB
project JCB/2021/000030 and Arnab Sen for discussions.

\appendix

\section{Computation of $\ham_F^{(2)}$ for cosine protocol}
\label{appendix}

The second order Floquet Hamiltonian can be broken into two parts
$\ham_F^{(2)} =  \ham_F^{(2a)} + \ham_F^{(2b)}$.
From Eq.~\eqref{eq:HF-2a} we get
\begin{align}
\label{app:HF-2a}
\ham_F^{(2a)} &= \frac{-i J^2}{2 \hbar T} \sum_{i, j} \sum_{m, n} B_{m, n}
\left[ J_m(\lambda \hat{A}_i) \cda_i c_{i+1} + J_m(-\lambda \hat{A}_i)  \right.
\nonumber\\
&\times \left. \cda_{i+1} c_i
\, , \, 
 J_n(\lambda \hat{A}_j) \cda_j c_{j+1} + J_n(-\lambda \hat{A}_j) \cda_{j+1} c_j \right],
\end{align}
where
\beq
\label{app:Bmn}
B_{m, n} \equiv \int_0^T d \tau_1 \int_0^{\tau_1} d \tau_2 \, e^{i m \omega \tau_1} 
e^{i n \om \tau_2}
\eeq
The evaluation of the above integrals yield
\begin{align}
\label{app:Bmn2}
B_{m, n} &= \frac{T^2}{2} \delta_{m,0} \delta_{n,0} + \frac{T}{i m \om} \delta_{n, 0}
(1 - \delta_{m, 0})
\nonumber\\
& - \frac{T}{i n \om} \delta_{m, 0} (1 - \delta_{n, 0})
+ \frac{T}{i n \om} (1 - \delta_{n, 0}) \delta_{m, -n}.
\end{align}
Due to the commutator structure of Eq.~\eqref{app:HF-2a} the first and the fourth terms above
do not contribute. The second and the third terms are non-zero and equal. Next, due to the 
$1/m$ factor in the second term, only integers $m$ contribute. We get
\begin{align}
\ham_F^{(2a)} &= - \frac{2J^2}{\hbar \om} \sum_{k=0}^{\infty} \frac{1}{2k +1}
\left[ \sum_i J_{2k+1} (\lambda \hat{A}_i) \right.
\nonumber\\
&\times \left.  \left( \cda_i c_{i+1} - {\rm h.c.} \right)
\, , \, \sum_j J_0 (\lambda \hat{A}_j) \left( \cda_j c_{j+1} + {\rm h.c.} \right) \right].
\end{align}
Noting that $\hat{A}_i$ can only take values 0, 1, -1, we have the relation
\[
J_{2k+1} (\lambda \hat{A}_i) = \hat{A}_i J_{2k+1} (\lambda).
\]
Using the above we get
\begin{align}
\label{app:HF-2a-2}
\ham_F^{(2a)} &= - \frac{2 J^2}{\hbar \om} \mathcal{C}(\lambda) 
\left[ \sum_i \hat{A}_i \left( \cda_i c_{i+1} - {\rm h.c.} \right) \right.
\,  , \, 
\nonumber\\
&\left. \sum_j J_0(\lambda \hat{A}_j) \left( \cda_j c_{j+1} + {\rm h.c.} \right) \right],
\end{align}
where
\[
\mathcal{C}(\lambda) \equiv \sum_{k = 0}^{\infty} \frac{J_{2k +1}(\lambda)}{2k +1}.
\]

For the second term $\ham_F^{(2b)}$ we note that
\[
\int_0^T d \tau_1 \int_0^{\tau_1} d \tau_2 \tilde{\ham}_p(\tau_2)
= \int_0^T d \tau_1 \int_{\tau_1}^T d \tau_2 \tilde{\ham}_p(\tau_1).
\]
Using the above relation and Eq.~\eqref{eq:HF-2b} we get
\begin{align}
\ham_F^{(2b)} &= \frac{i}{2 \hbar T} \int_0^T d \tau (T - 2\tau) \left[ \tilde{\ham}_p(\tau)
\, , \, \hat{K} \right]
\nonumber\\
&= \frac{-i J}{2 \hbar T}  \sum_{i, m} \int_0^T d \tau (T - 2\tau) e^{i m \om \tau}
\nonumber\\
&\times
\left[ J_m(\lambda \hat{A}_i) \cda_i c_{i+1} + J_m(-\lambda \hat{A}_i) \cda_{i+1} c_i
\, , \, \hat{K} \right]
\nonumber
\end{align}
For the $\tau$-integral above we use the relation
\[
 \int_0^T d \tau (T - 2\tau) e^{i m \om \tau} = - \frac{2T}{i m \om} (1- \delta_{m, 0}).
 \]
 The appearance of the factor $1/m$ in the above implies that, again, only the Bessel functions
 with odd indices contribute. This gives
 \beq
 \label{app:HF-2b}
 \ham_F^{(2b)} =  \frac{2 J \mathcal{C}(\lambda) }{\hbar \om} 
\left[ \sum_i \hat{A}_i \left( \cda_i c_{i+1} - {\rm h.c.} \right)
\, , \, \hat{K} \right].
\eeq
The Eqs.~\eqref{app:HF-2a-2} and \eqref{app:HF-2b} imply Eq.~\eqref{eq:HF-2-2}
in the main text.

\section{4-state ansatz for the long-time OTOC oscillations}
\label{otocobclong}

\begin{figure}
    \centering
    \includegraphics[width=1.0\linewidth]{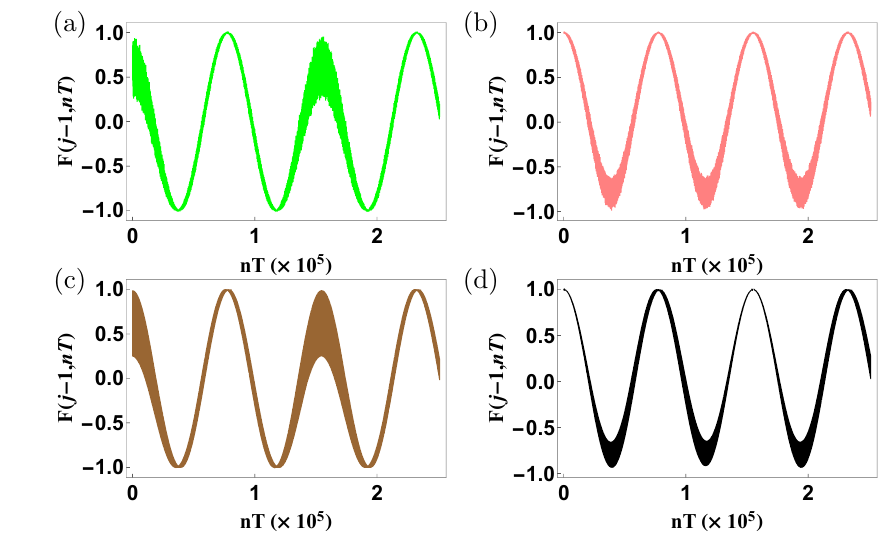}
    \caption{(Color Online) Comparison of exact results for OTOC $F(r_1,nT)$ with that obtained from Eq. \ref{F1F14} and \ref{f1f2detail}, starting from a $|\mathbb Z_2\rangle$ state. (a) Green (light-colored) line is obtained from exact dynamics for $j=1$,
    (b) Pink (light-colored) line is obtained from exact dynamics for $j=14$,
    (c) Brown (dark-colored) is obtained from the 4 state ansatz, given by Eq. \ref{estates} for $j=1$. (d)Black (dark-colored) is obtained from the 4 state ansatz, for $j=14$.  For all the plots, $L=14, V_0=2, V_2=6$ and $V_1=50$ with $\omega_D=V_1/2\hbar$.}
    \label{fig7}
\end{figure}
In this section, we discuss a simplification of Eq. \ref{otoc3} when a high value of the next-nearest-neighbor interaction $V_2$ is considered. In this case, the $\{|\mathbb Z_2\rangle,|\Bar{\mathbb Z }_2\rangle \}$ states, which have one extra next-nearest-neighbor pair, are well separated in energy from the other $|j_h\rangle$ states. Thus, the extent of hybridization between $\{|\mathbb Z_2\rangle,|\Bar{\mathbb Z }_2\rangle\}$ states and the $|j_h\rangle$ is small and hence the eigenfunctions of $H_F^{(1)}$ can be written down in terms of only a few Fock states as we show below.

We assume that the two highest energy levels $|\phi_7\rangle$ and $|\phi_8\rangle$ are mostly symmetric and anti-symmetric combinations of $\mathbb Z_2$ and $\Bar{\mathbb Z}_2$ with very small contributions from the two ``nearest" $|j_h\rangle$ states, i.e. $|j_h=2\rangle$ and $|j_h=12\rangle$. By ``nearest", we refer to states which can be connected to $\mathbb Z_2, \Bar{\mathbb Z}_2$ by one constrained hop. We also consider two more states $|\phi_{5,6}\rangle$ which are orthogonal to $|\phi_{7,8}\rangle$ and have energies $\epsilon_{5,6}$. We assume these last two states to be nearly degenerate i.e. $\epsilon_5\approx\epsilon_6=E_0$ and the splitting between the two highest states $\epsilon_8-\epsilon_7=\Omega\ll V_2$, $\epsilon_7-\epsilon_6=V_2$. Thus
\begin{align}
    |\phi_8\rangle &=& \frac{1}{\sqrt{2}}\mathcal C (|\mathbb Z_2\rangle-|\Bar{\mathbb Z}_2\rangle)+\frac{1}{\sqrt{2}}\mathcal S (|2\rangle-|12\rangle)\nonumber\\
    |\phi_7\rangle &=& \frac{1}{\sqrt{2}}\mathcal C (|\mathbb Z_2\rangle+|\Bar{\mathbb Z}_2\rangle)+\frac{1}{\sqrt{2}}\mathcal S (|2\rangle+|12\rangle)\nonumber\\
    |\phi_6\rangle &=& \frac{1}{\sqrt{2}}\mathcal S (|\mathbb Z_2\rangle-|\Bar{\mathbb Z}_2\rangle)-\frac{1}{\sqrt{2}}\mathcal C (|2\rangle-|12\rangle)\nonumber\\
    |\phi_5\rangle &=& \frac{1}{\sqrt{2}}\mathcal S (|\mathbb Z_2\rangle+|\Bar{\mathbb Z}_2\rangle)-\frac{1}{\sqrt{2}}\mathcal C (|2\rangle+|12\rangle)
    \label{estates}
\end{align}
where $\mathcal{C}=\cos{\theta}$ and $\mathcal{S}=\sin{\theta}$ with $\theta$ being a phenomenological parameter to be determined from diagonalization. Inverting these relations, the time evolved states read
\begin{eqnarray}
    |\mathbb Z_2(t)\rangle &=&\frac{1}{\sqrt{2}}\left[\mathcal C e^{-iV_2t} (|\phi_7\rangle + |\phi_8\rangle e^{-i \Omega t}) \right.\nonumber\\
    && \left.+ \mathcal{S}(|\phi_5\rangle+|\phi_6\rangle)\right]\nonumber\\
    |\Bar{\mathbb Z}_2(t)\rangle&=&\frac{1}{\sqrt{2}}\left[\mathcal C e^{-iV_2t} (|\phi_7\rangle \right.\nonumber\\
    && \left. - |\phi_8\rangle e^{-i \Omega t})+ \mathcal{S}(|\phi_5\rangle-|\phi_6\rangle)\right]\nonumber\\
    |2(t)\rangle&=&\frac{1}{\sqrt{2}}\left[\mathcal S e^{-iV_2t} (|\phi_7\rangle \right.\nonumber\\
    && \left. + |\phi_8\rangle e^{-i \Omega t})- \mathcal{C}(|\phi_5\rangle+|\phi_6\rangle)\right]\nonumber\\
    |12(t)\rangle&=&\frac{1}{\sqrt{2}}\left[\mathcal S e^{-iV_2t} (|\phi_7\rangle \right.\nonumber\\
    && \left. - |\phi_8\rangle e^{-i \Omega t})- \mathcal{C}(|\phi_5\rangle-|\phi_6\rangle)\right]
    \label{tevolstates}
\end{eqnarray}
where we have set the reference energy $E_0=0$. Using these 4 states, Eq. \ref{otoc2} for sites $j=1,14$ can be simplified as
\begin{eqnarray}
    F(0,nT)= f_1(nT) - f_2 (nT)\nonumber\\
    F(13,nT)=f_1(nT) + f_2 (nT)
    \label{F1F14}
\end{eqnarray}
where
\begin{eqnarray}
f_1(nT) &=& |\langle \mathbb Z_2(nT)|\Tilde{n}_1|\mathbb Z_2(nT)\rangle|^2 - |\langle \Bar{\mathbb Z}_2(nT)|\Tilde{n}_1|\mathbb Z_2(nT)\rangle|^2 \nonumber\\
    f_2(nT) &=& |\langle 2(nT)|\Tilde{n}_1|\mathbb Z_2(nT)\rangle|^2 + |\langle 12(nT)|\Tilde{n}_1|\mathbb Z_2(nT)\rangle|^2 \nonumber\\
    \label{f1f2}
\end{eqnarray}
Using the time-evolved states in Eq. \ref{tevolstates}, we obtain
\begin{widetext}
\begin{eqnarray}
    f_1(t)&=&\mathcal{C}^8 \cos{2\Omega t}-2 \mathcal{C}^6\mathcal{S}^2\sin{\Omega t}\big(\sin{(V_2+\Omega)t}-\sin{V_2t}\big)+2\mathcal{C}^4\mathcal{S}^4\left[2\cos{\Omega t}-1+2\big(1-\cos{(V_2+\Omega)t}-\cos{V_2t}\big)^2\right]\nonumber\\
    &-& 4 \mathcal{S}^2\mathcal{C}^2 \left(\mathcal{S}^4+\mathcal{C}^4\cos{\Omega t}\right)\left(1-\cos{(V_2+\Omega)t}-\cos{V_2t}\right)+\mathcal{S}^8\nonumber\\
    f_2(t)&=& \mathcal{S}^2\mathcal{C}^2 \left[\mathcal{S}^4 \big(\cos{(V_2+\Omega)t}-\cos{V_2t}\big)^2+\left(\mathcal{C}^2\sin{\Omega t}+\mathcal{S}^2(\sin{(V_2+\Omega)t}-\sin{V_2t})\right)^2+\left(\sin{(V_2+\Omega)t}+\sin{V_2t}\right)^2\right]\nonumber\\
    &+&\mathcal{S}^2\mathcal{C}^2 \left[\mathcal{S}^2 -\mathcal{C}^2\cos{\Omega t}+(2\mathcal{C}^2-1)\left(\cos{(V_2+\Omega)t}+\cos{V_2t}-1\right)\right]^2
    \label{f1f2detail}
\end{eqnarray}
\end{widetext}
For $V_2\gg J$, $\theta\ll 1$, we retain up to quadratic terms in $\theta$, yielding
\begin{eqnarray}
    F(0,t)&=&\left(1-4\theta^2\right)\cos{2\Omega t}+6\theta^2-2\theta^2\cos{\Omega t}\nonumber\\
    &+& 4\theta^2\cos{V_2 t}(\cos{\Omega t}-1)\nonumber\\
    F(13,t)&=&\left(1-4\theta^2\right)\cos{2\Omega t}-6\theta^2-6\theta^2\cos{\Omega t}\nonumber\\
    &+& 4\theta^2\cos{V_2 t}(3\cos{\Omega t}+1)
    \label{approxF1F2}
\end{eqnarray}
We compare in Fig. \ref{fig7} the exact results and those obtained from Eq. \ref{f1f2detail} for sites $j=1$ and $j=14$ where we find good agreement between the two. The parameters chosen are $V_0=2, V_2=6, V_1=50$ and $\omega_D=V_1/2\hbar$.

\section{OTOC in $\mathbb Z_2$ state with PBC}
\label{pbcotocz2}

\begin{figure}
    \centering
    \includegraphics[width=1.0\linewidth]{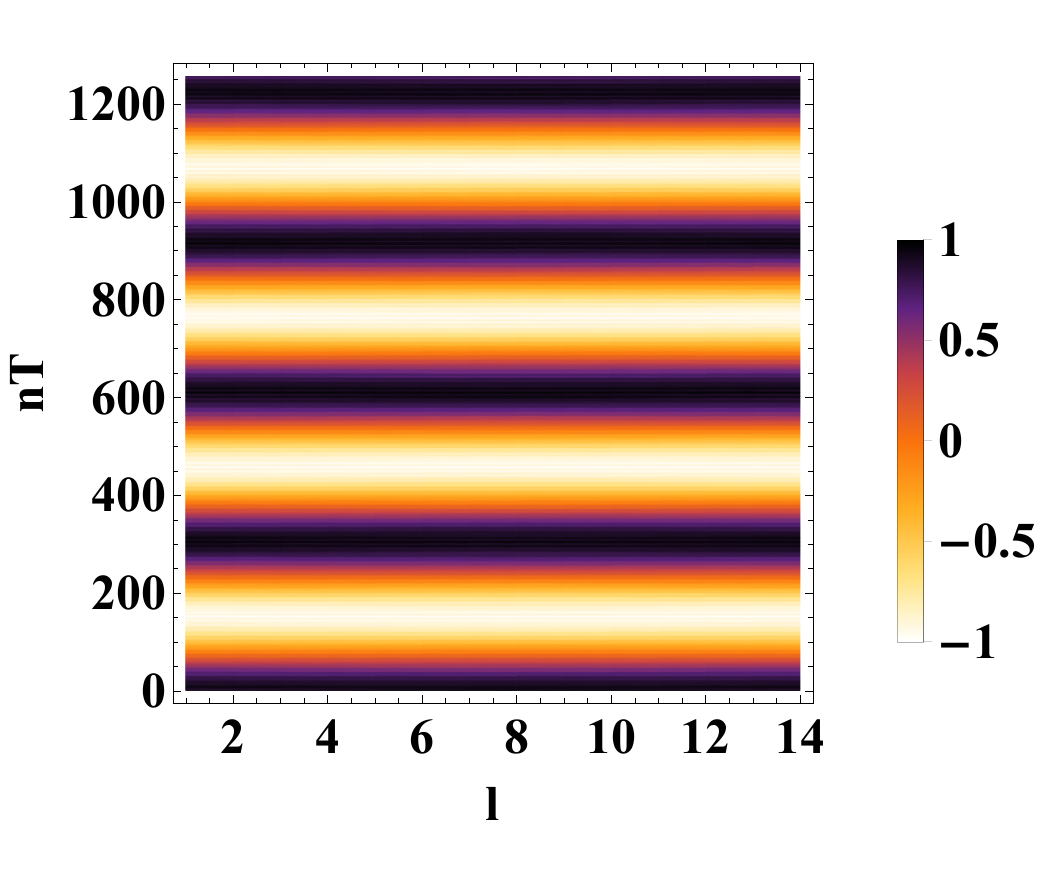}
    \caption{(Color Online) Profile of the OTOC $F(r,nT)$ in a $\mathbb Z_2$ state for $L=14$ with PBC and $\omega_D=V_1/2\hbar$. The initial operator is localized at $i=1$. This too shows alternate scrambling and unscrambling with a period $\tau\sim 300$. The parameters are $V_0=10V_2=2, V_1=20$}
    \label{fig8}
\end{figure}

In this appendix, we consider the evolution of the profile of the OTOC starting from a $|\mathbb Z_2\rangle$ state with PBC at the special frequency. Fig. \ref{fig8} shows that in this case too, there is alternate scrambling and unscrambling for $nT \leq 1200$ and $V_0=10 V_2=2, V_1=20$. In the following we show that 
these oscillations can be interpreted as tunneling back and forth between the states 
$\mathbb{Z}_2$ and $\bar{\mathbb{Z}}_2$ when the system is close to the fragmented limit. Such tunneling events were reported in Ref.~\cite{ourprl}. This can be understood by noting that for PBC, the $|\mathbb Z_2\rangle$ and $|\Bar{\mathbb Z}_2\rangle$ states are degenerate frozen states for $H_F^{(1)}$. This degeneracy is lifted by higher-order hopping terms and for exact $H_F$, there are two eigenstates which are symmetric and anti-symmetric combinations of $|\mathbb Z_2\rangle$ and $|\Bar{\mathbb Z}_2\rangle$ states viz
\begin{equation*}
    \chi_+ = \frac{1}{\sqrt{2}} (|\mathbb Z_2\rangle + |\Bar{\mathbb Z}_2\rangle), \quad \chi_-=\frac{1}{\sqrt{2}}(|\mathbb Z_2\rangle - |\Bar{\mathbb Z}_2\rangle)
\end{equation*}.
The energy splitting between these two states is given by $\Omega=\epsilon_- - \epsilon_+$. This implies the time evolutions
\begin{align}
| \mathbb{Z}_2(t)\ra &= e^{i \epsilon_- t} \left[ (e^{i\Omega t} + 1) | \mathbb{Z}_2 \ra
+ (e^{i\Omega t} - 1) |  \bar{\mathbb{Z}}_2 \ra \right]/2,
\nonumber \\
|  \bar{\mathbb{Z}}_2(t) \ra &= e^{i \epsilon_- t} \left[ (e^{i\Omega t} - 1) | \mathbb{Z}_2 \ra
+ (e^{i\Omega t} + 1) |  \bar{\mathbb{Z}}_2 \ra \right]/2.
\label{z2tz2bart}
\end{align} 
Inserting an approximate complete set comprising of the states $\mathbb Z_2$ and $\bar{\mathbb Z}_2$ in Eq. \ref{otoc1} of the main text, we obtain
\begin{align}
\label{eq:long-osc1}
&F(r_1,t)  \approx \left| \la \mathbb{Z}_2 | \Tilde{n}_1(t) |  \mathbb{Z}_2 \ra \right|^2
- \left| \la \mathbb{Z}_2 | \Tilde{n}_1(t) |  \bar{\mathbb{Z}}_2 \ra \right|^2
\nonumber \\
&= \left| \la \mathbb{Z}_2(t) | \Tilde{n}_1 |  \mathbb{Z}_2(t) \ra \right|^2
- \left| \la \mathbb{Z}_2(t) | \Tilde{n}_1 |  \bar{\mathbb{Z}}_2(t) \ra \right|^2,
\end{align}
where the last line is going from Heisenberg to Schrodinger picture. Note, the above equation already captures an important aspect of Fig.\ \ref{fig8}, namely $F(r_1,t)$ is mostly 
independent of $r_1$ at this timescale.

Using Eq. \ref{z2tz2bart},we get
\beq
F(r_1,t) \approx \cos (2 \Omega t).
\eeq 
Thus, at stroboscopic times $t = nT$, where $n\Omega T$ is close to a integer multiple of $\pi$, 
the OTOC is close to one while, when $n\Omega T$ is close to a half-integer mutiple of $\pi$,
the OTOC is close to minus one.


\end{document}